\documentclass[aps,prd,notitlepage,showpacs,nofootinbib,superscriptaddress]{revtex4-1}

\usepackage{graphicx}
\usepackage[utf8]{inputenc} 

\usepackage{amsmath}
\usepackage{amssymb}
\usepackage{float}
\usepackage{comment}
\usepackage{slashed}
\usepackage[normalem]{ulem}

\usepackage{booktabs}

\usepackage{mathtools,braket,blkarray}

\usepackage{amsmath}
\usepackage{amssymb}

\usepackage[usenames,dvipsnames]{color}
\usepackage[colorinlistoftodos]{todonotes}
\usepackage[colorlinks=true,citecolor=darkred,urlcolor=darkred, pdfborder={0 0 0}]{hyperref}
\usepackage[normalem]{ulem}

\usepackage{multirow}
\definecolor{darkred}{rgb}{0.6,0,0}
\definecolor{darkgreen}{rgb}{0.0, 0.4, 0.0}
\usepackage[colorinlistoftodos]{todonotes}

\definecolor{linkcolor}{rgb}{0,0,0.5}

\newcommand{\jv}[1]{{\color{red}  #1}}

\newcommand{\cv}[1]{{\color{darkgreen}  #1}}
\newcommand {\ignore}[1]{}

\def \znbb {$\rm 0\nu\beta\beta$ }

\def\gsim{\raise0.3ex\hbox{$\;>$\kern-0.75em\raise-1.1ex\hbox{$\sim\;$}}}
\def\lsim{\raise0.3ex\hbox{$\;<$\kern-0.75em\raise-1.1ex\hbox{$\sim\;$}}}

\def\SM{$\mathrm{SU(3)_c \otimes SU(2)_L \otimes U(1)_Y}$ }

\usepackage{soul}

\definecolor{mightnightblue}{RGB}{25,25,112}

\definecolor{brown}{rgb}{0.59, 0.29, 0.0}

\def\vev#1{\left\langle #1\right\rangle}
\def\SM{$\mathrm{SU(3)_c \otimes SU(2)_L \otimes U(1)_Y}$ }
\def\21{$\mathrm{SU(2)_L \otimes U(1)_Y}$}

\newcommand{\AddrAHEP}{%
  AHEP Group, Institut de F\'{i}sica Corpuscular --
  C.S.I.C./Universitat de Val\`{e}ncia, Parc Cient\'ific de Paterna.\\
 C/ Catedr\'atico Jos\'e Beltr\'an, 2 E-46980 Paterna (Valencia) - SPAIN}

\begin{document}


\title{\boldmath \color{BrickRed} Flavour and CP predictions from orbifold compactification}

\author{Francisco J. de Anda}\email{fran@tepaits.mx}
\affiliation{Tepatitl{\'a}n's Institute for Theoretical Studies, C.P. 47600, Jalisco, M{\'e}xico}

\author{Jos\'{e} W. F. Valle}\email{valle@ific.uv.es}
\affiliation{\AddrAHEP}

\author{Carlos A. Vaquera-Araujo}\email{vaquera@fisica.ugto.mx}
\affiliation{Consejo Nacional de Ciencia y Tecnolog\'ia, Av. Insurgentes Sur 1582. Colonia Cr\'edito Constructor, Del. Benito Ju\'arez, C.P. 03940, Ciudad de M\'exico, M\'exico}
\affiliation{Departamento de F\'isica, DCI, Campus Le\'on, Universidad de
Guanajuato, Loma del Bosque 103, Lomas del Campestre C.P. 37150, Le\'on, Guanajuato, M\'exico}

\begin{abstract}
\vspace{0.5cm}

We propose a theory for fermion masses and mixings in which an $A_4$  family symmetry arises naturally from a six-dimensional spacetime after orbifold compactification.
The flavour symmetry leads to the successful ``golden'' quark-lepton unification formula. 
The model reproduces oscillation parameters with good precision, giving sharp predictions for the CP violating phases of quarks and leptons,
in particular $\delta^\ell \simeq +268 ^\circ$. The effective neutrinoless double-beta decay mass parameter is also sharply predicted as $\vev{ m_{\beta\beta}} \simeq 2.65\,\mathrm{meV}$.
  
\end{abstract}

\maketitle
\noindent

\section{Introduction}
\label{Sect:intro}

The historical discovery of neutrino oscillations~\cite{McDonald:2016ixn,Kajita:2016cak} implies the need for neutrino masses on the one hand, and a way to understand the special pattern 
of lepton mixing angles on the other~\cite{deSalas:2017kay}.
Altogether, this has spurred interest in understanding the theoretical origin of neutrino mass and the understanding of flavour parameters. 

The Standard Model (SM) lacks neutrino masses as well as an organizing principle in terms of which to understand flavour.
Therefore the SM should be supplemented by the inclusion of finite neutrino masses as well as by some explanation of the mixing pattern. 
In order to explain the observed pattern of lepton mixing, one must therefore go beyond the simplest seesaw picture, in order to tackle the flavour problem. 
The peculiar pattern of lepton mixing parameters is unlikely to be accidental.
Family symmetries should be lurking out there, and somehow coupled with the seesaw mechanism. 
Unfortunately, unveiling the underlying family symmetry capable of reproducing the observed pattern of quark and lepton masses and mixings is a great challenge.
The symmetry tool-kit provided by mathematics is too wide~\cite{Ishimori:2010au}. 

A new direction to approach the flavour problem is the use of extra dimensions.
For example, theories with extra warped dimensions open the way for a geometrical description of mass hierarchies.
This constitutes a generalization of the idea of Randall and Sundrum whose aim was to address just the weak gauge hierarchy problem~\cite{Randall:1999ee}. 
It has been shown that such a geometrical view of fermion mass hierarchies can be made consistent with the imposition of family symmetries to account for fermion mixing~\cite{Chen:2015jta}. 

In this paper we propose to ``derive'' the family symmetry itself from the presence of flat extra dimensions~\cite{Antoniadis:1998ig}.  
We use the framework of 6-dimensional theories compactified on a torus~\cite{deAnda:2018oik,deAnda:2018yfp}. 
This way we obtain a predictive model for fermion masses and mixings in which the family symmetry $A_4$ emerges naturally 
as a remnant symmetry after orbifold compactification.
This provides an alternative and elegant realization of the $A_4$ family symmetry in terms of SM fields.

In contrast to the original $A_4$ proposal in \cite{Babu:2002dz}, the present realization also makes predictions for fermion mass hierarchies, not just angles and phases. 
In particular, it predicts the ``golden'' quark-lepton unification formula, which also emerges in other schemes~\cite{Morisi:2011pt,King:2013hj,Morisi:2013eca,Bonilla:2014xla}.
However, in those early models the family symmetry was imposed \emph{a priori}.  
Our new proposal will be tested in many ways, especially by the measurement of the leptonic CP phase in neutrino oscillations, 
and by an improved measurement of quark and lepton masses, posing a challenge for future neutrinoless double-beta decay experiments.

The paper is organised as follows. In Sec.~\ref{Sec:setup} we present the theoretical setup, while in Sec.~\ref{sec:model} we sketch the actual 
model construction, its field content and quantum numbers. In Sec.~\ref{sec:analysis} we give a numerical analysis of the resulting predictions and conclude 
in Sec.~\ref{sec:Conclusions}.

\section{Theoretical setup}
\label{Sec:setup}

In this paper we study a realistic model for quark and lepton masses and mixings, based on a discrete $A_4$ family symmetry that emerges naturally from an extra dimensional framework.
In this section we outline how $A_4$ can be obtained as a remnant symmetry from the orbifold compactification of a 6 dimesional spacetime. 

We assume the spacetime manifold as $\mathcal{M}=\mathbb{M}^4\times (\mathbb{T}^2/\mathbb{Z}_2)$, where the torus $\mathbb{T}^2$ is defined by the relations
\begin{equation}\begin{split}
(x^5,x^6)&=(x^5+2\pi R_1,x^6),
\\ (x^5,x^6)&=(x^5+2\pi R_2\cos\theta,x^6+2\pi R_2\sin\theta),
\label{eq:tras}
\end{split}\end{equation}
and the radii of the extra dimensions are taken to be of the same order of magnitude.
This allows us to define the compactification scale $R_1\sim R_2\sim 1/M_C$.  
In the following, we will focus on the particular choice for the twist angle $\theta=2\pi/3$, and we will show how this specific angle can lead to the emergence of a remnant discrete symmmetry.
The $\mathbb{Z}_2$ orbifolding of the torus comes from the identification 
\begin{equation}
(x^5,x^6)=(-x^5,-x^6).
\label{eq:par}
\end{equation}
In order to simplify the analysis it is convenient to rescale the original radii of the torus as $2\pi R_1\Rightarrow 1$ and $2\pi R_2\Rightarrow 1$ and to adopt the complex notation $z=x_5+ix_6$ 
With the help of the root of unity  $\omega=e^{i\theta}=e^{2i\pi/3}$, the symmetries of the orbifold from Eqs.(\ref{eq:tras},\ref{eq:par}) can be written as 
\begin{equation}\begin{split}
z&=z+1,\\
z&=z+\omega,\\
z&=-z.
\label{eq:orbtra}
\end{split}\end{equation}
There are 4 fixed points under these orbifold transformations, that define four invariant 4-dimensional branes 
\begin{equation}
\bar{z}=\left\{0,\ \frac{1}{2},\ \frac{\omega}{2},\ \frac{1+\omega}{2}\right\}.
\label{eq:orgbro}\end{equation}
There is an additional symmetry of the set of branes, inherited from the Poincar\'e invariance of the extra dimensional part of the manifold.
After orbifold compactification, the remaining transformations that permute the four branes leaving the whole brane set invariant are 
\begin{equation}
S_1:z\to z+1/2,\ \ \ S_2:z+\omega/2,\ \ \ R:z\to\omega^2 z,
\label{eq:remsym}
\end{equation}
which are just translations and rotations. Therefore, the fields localized on the branes must transform accordingly to this remnant symmetry 
\cite{Altarelli:2006kg,Adulpravitchai:2009id,Adulpravitchai:2010na}. We can write these transformations explicitly as 
\begin{equation}
S_1=[(12)(34)],\ S_2=[(13)(24)],\ R=[(1)(243)].
\end{equation}
Among this set of transformations there are only two independent ones, since $S_2=R^2\cdot S_1\cdot R$. 
These symmetry transformations relate to the $A_4$ generators through the identification $S=S_1,\ T=R,$ satisfying  
\begin{equation}
S^2=T^3=(ST)^3=1,
\end{equation}
which is the presentation for the $A_4$ group. 

From the above prescription, we can build models based on a remnant symmetry $A_4$, or more generally $S_4$, given the fact that the fields localized on the branes can transform as a $\mathbf{4}$ of the permutation group $S_4$. 
In this work, we adopt the minimalist choice of a remnant $A_4$ flavour group, and a specific embedding such that the representation $\mathbf{4}$ of $S_4$ decomposes into irreducible representations of $A_4$ as $\mathbf{4}\to\mathbf{3}+\mathbf{1}$~\cite{Bazzocchi:2009pv,deAnda:2018oik}.
Henceforth we can single out the brane fields as transforming into any of the two irreps $\mathbf{3}$ or $\mathbf{1}$ of $A_4$. 

In summary, in this framework, the flavour symmetry naturally emerges from the branes located at the fixed points of the spacetime manifold.
We can generalize this statement by the following requirement: As the fields located at the branes experience a subgroup of the extra dimensional part of the Poincaré group,
then all 6-dimensional fields should also transform under some irreducible representation of the $A_4$ remnant symmetry \cite{Burrows:2009pi}.

\section{Simplest Model}
\label{sec:model}

Our model features a six-dimensional implementation of the standard \SM gauge symmetry including 3 ``right-handed'' neutrinos and supplemented with the orbifold
compactification described in the previous section.
The transformation properties of the fields under the remnant $A_4$ symmetry and their localization on the orbifold are shown in Table \ref{tab:fields}. 
Notice that all fields, except the right-handed quarks, transform as flavour triplets. 
\begin{table}[h]
\centering
\footnotesize
\begin{tabular}{ |c|ccc|c|}
\hline
\textbf{Field} &$\qquad$ & $A_4$ &$\qquad$& Localization \\
\hline
$L$ & & $\mathbf{3}$  &  &Brane\\
$d^c$ && $\mathbf{3}$  && Brane\\
$e^c$ && $\mathbf{3}$  && Brane\\
$Q$ && $\mathbf{3}$  &&Brane\\
$u_{1,2,3}^c$ && $\mathbf{1''},\mathbf{1'},\mathbf{1}$ &&Bulk \\
$\nu^c$ && $\mathbf{3}$ &&Brane \\
\hline
$H_u$ && $\mathbf{3}$ && Brane \\
$H_\nu$ && $\mathbf{3}$ & & Brane \\
$H_d$  && $\mathbf{3}$  &&Brane\\
$\sigma$ && $\mathbf{3}$ && Bulk\\
\hline
\end{tabular}
\caption{Field content of the model.}
\label{tab:fields}
\end{table}

The scalar sector consists of 3 Higgs doublets, which are flavour triplets, and an extra singlet scalar $\sigma$ driving spontaneous breaking of lepton number at the high energy regime. 
We have one Higgs doublet for the down-type quarks and charged leptons, and two Higgs doublets that give the up-type quark masses and Dirac masses to neutrinos. 
Notice that $H_d$ only couples to down type fermions (charged leptons and down quarks), while $H_u$ couples only to up quarks and $H_\nu$ only couples to neutrinos. 
The effective Yukawa terms after compactification are 
\begin{equation}
\begin{split}
\mathcal{L}_Y &= y^N  \nu^c \nu^c \sigma+y_1^\nu(L H_\nu \nu^c)_1+y_2^\nu(L H_\nu \nu^c)_2\\
 &\quad +y_1^d(Q d^c H_d)_1+y_2^d(Q d^c H_d)_2+y_1^e(L e^c H_d)_1+y_2^e(L e^c H_d)_2\\
&\quad +y_1^u(QH_u)_{1'}u_{1}^c+y_2^u(QH_u)_{1''}u_{2}^c+y_3^u(QH_u)_{1}u_{3}^c,
\label{eq:yuk}
\end{split}
\end{equation}  
where the notation $()_{1,2}$ stands for the possible singlet contractions $\mathbf{3}\times \mathbf{3} \times \mathbf{3}\to \mathbf{1}_{1,2}$ and
$\mathbf{3}\times \mathbf{3} \to \mathbf{1}_{1,1',1''}$ in $A_4$. In this work, we assume CP symmetry to be respected at high energies, and impose a non-trivial CP symmetry that does not commute with the $A_4$ family symmetry \begin{equation}
\mathcal{CP}:\ \ \phi_{\textbf{1}}\to\phi_{\textbf{1}}^\dagger,\ \ \phi_{\textbf{1}'}\to\phi_{\textbf{1}''}^\dagger,\ \ \phi_{\textbf{1}''}\to \phi_{\textbf{1}'}^\dagger,\ \ \phi_{\textbf{3}}\to \left(\begin{array}{ccc}1&0&0\\0&0&1\\0&1&0\end{array}\right) \phi_{\textbf{3}}^\dagger,
\end{equation} 
which is a consistent generalized CP symmetry \cite{Ding:2013bpa,Chen:2015siy}. We work in the Ma-Rajasekaran basis \cite{Ma:2001dn,Barry:2010zk}, where this generalized CP conservation implies all Yukawa couplings in Eq. \ref{eq:yuk} to be real (see Appendix \ref{app:a4} for further details).

The scalar field $\sigma$ gets a vacuum expectation value (VEV) that breaks spontaneously $U(1)_L$ and $A_4$, giving high-scale masses to the ``right-handed'' neutrinos, 
and playing the dual role of a Majoron and a (renormalizable) flavon~\cite{Gelmini:1983ea}. Since this field propagates in the bulk of the orbifold, it must satisfy the boundary condition 
\begin{equation}
\sigma(x,z)=P \sigma(x,-z),
\end{equation}
where $P$ is an arbitrary symmetry transformation matrix such that $P^2=1$. After spontaneous symmetry breaking (SSB), the corresponding VEV is subject to the boundary condition
\begin{equation}
\braket{\sigma}=P \braket{\sigma}.
\end{equation}
We choose the matrix $P$ to be trivial for all singlets, except for the field $\sigma$, which is the only flavour triplet in the bulk. In order to identify the non-rivial $P$ transformation of $\sigma$ we observe that there is only one term in the bulk involving this field and it is its kinetic term. This is the only term affected by $P$ and it is invariant under an $SU(3)$ symmetry transformation. Therefore the matrix $P$ is chosen arbitrarily with the conditions $P\in SU(3)$ and $P^2=1$ for consistency. Staring from a phenomenologically viable $\braket{\sigma}$ alignment, we can choose a suitable $P$
\begin{equation}\label{sigmavev}
P=\frac{1}{3}\left(\begin{array}{ccc}-1&2\omega^2& 2\omega \\ 2\omega & -1& 2\omega^2\\ 2\omega&2\omega&-1\end{array}\right)\ \ \ \Longrightarrow\ \ \ 
\braket{\sigma}=v_\sigma\left(\begin{array}{c}1\\ \omega\\ \omega^2\end{array}\right),
\end{equation}
with $\omega=e^{2\pi i/3}$. Therefore the VEV can be aligned via the boundary condition without the need of an alignment potential \cite{deAnda:2018yfp}. This VEV preserves the generalized CP but its phases become physical after the $A_4$ breaking in the Yukawa sector \cite{Ivanov:2014doa,deMedeirosVarzielas:2017ote}. The structure of the phases in that alignment is completely determined by the $A_4$ breaking boundary condition. 

Since the $A_4$ symmetry is broken at a high energy scale, the Higgs doublets obtain the most general $A_4$ breaking VEVs, that we assume to be real, breaking the generalized CP symmetry involving $A_4$ but preserving trivial CP. We parametrize them as 
\begin{equation}\label{vevs}\begin{split}
\braket{H_u}=v_u\left(\begin{array}{c}\epsilon_1^u\\ \epsilon_2^u \\ 1\end{array}\right),\ \ \ \ \braket{H_\nu}=v_\nu \left(\begin{array}{c}\epsilon_1^\nu \\ \epsilon_2^\nu \\ 1\end{array}\right),\ \ \ \braket{H_d}=v_d \left(\begin{array}{c}\epsilon_1^d \\ \epsilon_2^d \\ 1\end{array}\right),
\end{split}\end{equation}
with real parameters $v_u,v_\nu,v_d,\epsilon^{u,\nu,d}_{1,2}$. After spontaneous symmetry breaking (SSB) the mass matrices of quarks and leptons become 
\begin{equation}\begin{split}
 M_u&=v_u\left(\begin{array}{ccc} y_1^u\epsilon_1^u &y_2^u \epsilon_1^u & y_3^u\epsilon_1^u \\
 y_1^u\epsilon_2^u \omega^2&  y_2^u\epsilon_2^u \omega &   y_3^u\epsilon^u_2 \\
  y_1^u \omega& y_2^u \omega^2&y_3^u\end{array}\right),\\
M_d&=v_d\left(\begin{array}{ccc} 0 & y_1^d\epsilon_1^d & y_2^d \epsilon_2^d \\
 y_2^d\epsilon_1^d & 0 &  y_1^d\\
 y_1^d\epsilon_2^d & y_2^d&0\end{array}\right),\\
M_e&=v_d\left(\begin{array}{ccc} 0 & y_1^e\epsilon_1^d & y_2^e \epsilon_2^d \\
 y_2^e\epsilon_1^d & 0 &  y_1^e\\
 y_1^e\epsilon_2^d & y_2^e&0\end{array}\right),\\
M_N^R&=y^N v_\sigma\left(\begin{array}{ccc}0 &   \omega^2 & \omega \\  \omega^2 & 0&1\\  \omega&1&0\end{array}\right),\\
M_\nu^D&=v_\nu\left(\begin{array}{ccc} 0 & y_1^\nu\epsilon_1^\nu & y_2^\nu \epsilon_2^\nu \\
 y_2^\nu\epsilon_1^\nu & 0 &  y_1^\nu\\
 y_1^\nu\epsilon_2^\nu & y_2^\nu&0\end{array}\right),\\[.2cm]
M_\nu^L&=M_\nu^D(M_N^R)^{-1}(M_\nu^D)^T,
\label{eq:massmat}
\end{split}\end{equation}
 where all Yukawa couplings are forced to be real due to the CP symmetry.

\section{ Flavour and CP predictions}
\label{sec:analysis}

\subsection{Quark and lepton mass hierarchies}
\label{sec:mass-hierarchies}

The first aspect of the flavour problem consists in explaining the fermion mass hierarchy. 
We start this section by pointing the first prediction of the model: the golden relation between charged lepton and down-type quark masses,
\begin{equation}\label{golden}
\frac{m_\tau}{\sqrt{m_\mu m_e}}\approx\frac{m_b}{\sqrt{m_s m_d}},
\end{equation}
that emerges from the fact that both the charged leptons and down quarks obtain their masses from the same Higgs doublet $H_d$.
This relation was already noted in different contexts~\cite{Morisi:2011pt,King:2013hj,Morisi:2013eca,Bonilla:2014xla}
in which the flavour symmetry was imposed by hand. And also in Ref.~\cite{Reig:2018ocz} in which the golden formula was related to a Peccei-Quinn symmetry.

Here the golden formula arises since the mass matrices share the same structure fixed by the remnant $A_4$ symmetry.
This relation is in good agreement with experiments and robust against renormalization group evolution. 
Although we are far from a complete and comprehensive theory of flavour, we think that Eq.(\ref{golden}) could well be part of the ultimate, yet-to-be-found theory. 

\subsection{Quark and lepton mixing patterns}
\label{sec:mixing-patterns}

Beyond the issue of fermion mass hierarchies, the flavour problem also includes the challenge of explaining the observed pattern of fermion mixing.
In particular, explaining the disparity between quark and lepton mixing angles. 
For our numerical analysis of the model we adopt the symmetrical presentation of fermion mixing given in~\cite{Schechter:1980gr}. 
\begin{equation}  \label{eq:CKM}
 V_{CKM}= \left( 
\begin{array}{ccc}
c^q_{12} c^q_{13} & s^q_{12} c^q_{13}  & s^q_{13} e^{-i\delta^q}
\\ 
-s^q_{12} c^q_{23}- c^q_{12} s^q_{13} s^q_{23} e^{ i \delta^q} & c^q_{12} c^q_{23} - s^q_{12} s^q_{13} s^q_{23} e^{ i \delta^q } & c^q_{13} s^q_{23} \\ 
s^q_{12} s^q_{23} - c^q_{12} s^q_{13} c^q_{23} e^{ i\delta^q } & - c^q_{12} s^q_{23}  - s^q_{12} s^q_{13} c^q_{23} e^{i\delta^q } & c^q_{13} c^q_{23}%
\end{array}
\right)\,, 
\end{equation}

For the case of the quarks this coincides with the Cabibbo-Kobayashi-Maskawa (CKM) matrix form used in the PDG~\cite{Tanabashi:2018oca},
while for the leptons the mixing matrix includes also the new phases associated to the Majorana nature of neutrinos~\cite{Schechter:1980gr}.
In the symmetrical presentation the lepton mixing matrix is given as, 
\begin{equation}  \label{eq:symmetric_para}
 K= \left( 
\begin{array}{ccc}
c^{\ell}_{12} c^{\ell}_{13} & s^{\ell}_{12} c^{\ell}_{13} e^{ - i \phi_{12} } & s^{\ell}_{13} e^{ -i \phi_{13} }
\\ 
-s^{\ell}_{12} c^{\ell}_{23} e^{ i \phi_{12} } - c^{\ell}_{12} s^{\ell}_{13} s^{\ell}_{23} e^{ -i ( \phi_{23} -
\phi_{13} ) } & c^{\ell}_{12} c^{\ell}_{23} - s^{\ell}_{12} s^{\ell}_{13} s^{\ell}_{23} e^{ -i ( \phi_{23} +
\phi_{12} - \phi_{13} ) } & c^{\ell}_{13} s^{\ell}_{23} e^{- i \phi_{23} } \\ 
s^{\ell}_{12} s^{\ell}_{23} e^{ i ( \phi_{23} + \phi_{12} ) } - c^{\ell}_{12} s^{\ell}_{13} c^{\ell}_{23} e^{ i
\phi_{13} } & - c^{\ell}_{12} s^{\ell}_{23} e^{ i \phi_{23} } - s^{\ell}_{12} s^{\ell}_{13} c^{\ell}_{23} e^{
-i ( \phi_{12} - \phi_{13} ) } & c^{\ell}_{13} c^{\ell}_{23}%
\end{array}
\right)\,,
\end{equation}
with $c^{f}_{ij}\equiv \cos\theta^f_{ij}$ and $s^{f}_{ij}\equiv \sin\theta^f_{ij}$.
The advantage of using the symmetrical parameterization resides in the transparent role played by the CP phases.

First of all we have a ``rephasing invariant'' form for the standard ``Dirac'' CP violating phase
\begin{equation}
\delta^{\ell}=\phi_{13}-\phi_{12}-\phi_{23},
\end{equation}
that enters in the description of neutrino oscillations.
In addition, we have a conceptually transparent description of the Majorana phases in which these, and only these, enter in the effective mass parameter characterizing the amplitude for neutrinoless double beta decay~\cite{Rodejohann:2011vc},
\begin{equation}
\langle m_{\beta\beta}\rangle=\left|\sum_{j=1}^3 K_{ej}^2 m_j\right|=
\left|c^{\ell\,2}_{12}c^{\ell\,2}_{13} m_1 + s^{\ell\,2}
_{12}c^{\ell\,2}_{13} m_2 e^{2i\phi_{12} }+ s^{\ell\,2}_{13} m_3 e^{2i\phi_{13}}\right|.
\end{equation}

The present model is characterized by 15 parameters ($y_{1,2}^{\nu}v_\nu\,,y_{1,2}^{e,d}v_d,\ y_{1,2,3}^uv_u,\ \epsilon_{1,2}^{u,\nu,d}$) that must reproduce the 22 flavour observables at low energies.
The latter include the 12 mass parameters i.e. $m_{u,c,t,d,s,b,e,\mu,\tau},\ m^{\nu}_{1,2,3}$ plus the 4 CKM mixing angles and phase $\theta_{12,13,23}^q,\ \delta^q$, 
the 3 neutrino mixing angles $ \theta^l_{12,13,23}$ and 3 CP phases $\phi_{12,13,23}$~\footnote{The extra parameter $y^Nv_\sigma$ setting the ``right-handed'' mass scale can be reabsorbed. }. \\[-.3cm]

There are thus 7 predictions in the model at low energies, among which we can readily identify the emergence of the golden relation in Eq.(\ref{golden}).
In addition, one has the four CP violating phases $\delta^q$, and $\phi_{12,13,23}$, that arise from the single fixed phase induced by the $ \braket{\sigma}$ alignment in Eq. (\ref{sigmavev}). 

\subsection{Global Flavour Fit}
\label{sec:global-flavour-fit}

In order to better identify the physics predictions, we perform a fit to determine the model parameters from the available experimental flavour data. 
The fit to the data is performed as follows: we define the chi-square function
\begin{equation}
\chi^2=\sum (\mu_{\text{exp}}-\mu_{\text{model}})^2/\sigma_{\text{exp}},
\end{equation}
where the sum runs through the 19 measured physical parameters. \\[-.2cm]

\begin{table}[ht]
	\centering
	\footnotesize
	\renewcommand{\arraystretch}{1.1}
	\begin{tabular}[t]{|lc|r|}
		\hline
		Parameter &\qquad& Value \\ 
		\hline
		$y^e_1v_d/\mathrm{GeV}$ &\quad& $1.745$ \\
		$y^e_2v_d/(10^{-1}\mathrm{GeV})$ && $1.019$ \\
		\rule{0pt}{3ex}%
		$y^d_1v_d/(10^{-2}\mathrm{GeV})$ &&$-4.690$ \\
		$y^d_2v_d /\mathrm{GeV}$ && $-2.914$ \\
		\rule{0pt}{3ex}%
		$y^\nu_{1} v_\nu/\sqrt{Y^N v_\sigma \mathrm{meV}}$ &&$-7.589$ \\
		$y^\nu_2v_\nu/(\sqrt{Y^N v_\sigma \mathrm{meV}}\times10^{-1} )$ && $-6.980$ \\
		\rule{0pt}{3ex}%
		$y^u_{1} v_u/(10^{-1}\mathrm{GeV})$ &&$-5.998$ \\
		$y^u_2v_u/(10^2\mathrm{GeV})$ && $1.712$ \\
		$y^u_3v_u/\mathrm{GeV}$ && $7.105$ \\
		\rule{0pt}{3ex}%
		$\epsilon^u_1/10^{-4}$ && $-7.213 $ \\
		$\epsilon^u_2/10^{-2}$ && $-5.080$ \\
		\rule{0pt}{3ex}%
		$\epsilon^d_1/10^{-3}$ && $-2.658$ \\
		$\epsilon^d_2/10^{-3}$ && $-6.199$ \\
		\rule{0pt}{3ex}%
		$\epsilon^\nu_1/10^{-1}$ && $1.691$ \\
		$\epsilon^\nu_2/10^{-1}$ && $8.542$ \\
		\hline	
	\end{tabular}
	\hspace*{0.5cm}
	\begin{tabular}[t]{ |l |c|c c |c| }
		\hline
		\multirow{2}{*}{Observable}& \multicolumn{2}{c}{Data} & & \multirow{2}{*}{Model best fit}  \\
		\cline{2-4}
		& Central value & 1$\sigma$ range  &   & \\
		\hline
		$\theta_{12}^\ell$ $/^\circ$ & 34.44 & 33.46 $\to$ 35.67 && $34.36$  \\ 
		$\theta_{13}^\ell$ $/^\circ$ & 8.45 & 8.31 $\to$ 8.61  && $8.31$  \\  
		$\theta_{23}^\ell$ $/^\circ$ & 47.69 & 45.97 $\to$ 48.85  && $48.47$ \\ 
		$\delta^\ell$ $/^\circ$ & 237 & 210 $\to$ 275 && $268$  \\
		$m_e$ $/ \mathrm{MeV}$ & 0.489 &  0.489 $\to$ 0.489 && $0.489$ \\ 
		$m_\mu$ $/  \mathrm{GeV}$ & 0.102 & 0.102  $\to$ 0.102  &&  $0.102$ \\ 
		$m_\tau$ $/ \mathrm{GeV}$ &1.745 & 1.743 $\to$1.747 && $1.745$ \\ 
		$\Delta m_{21}^2 / (10^{-5} \, \mathrm{eV}^2 ) $ & 7.55  & 7.39 $\to$ 7.75 && $7.63$  \\
		$\Delta m_{31}^2 / (10^{-3} \, \mathrm{eV}^2) $ & 2.50  & 2.47 $\to$ 2.53 &&  $2.42$ \\
		$m_1$ $/\mathrm{meV}$  & & & & $4.12$ \\ 
		$m_2$ $/\mathrm{meV}$  & && & $9.66$ \\ 
		$m_3$ $/\mathrm{meV}$  & && & $50.11$ \\
		$ \phi_{12} $ $/^\circ$ & & && $250$  \\
		$ \phi_{13}$  $/^\circ$& & && $187$  \\    
		$ \phi_{23}$  $/^\circ$& & && $29$  \\    	
		\hline
		$\theta_{12}^q$ $/^\circ$ &13.04 & 12.99 $\to$ 13.09 &&  $13.04$ \\	
		$\theta_{13}^q$ $/^\circ$ &0.20 & 0.19 $\to$ 0.22 && $0.20$  \\
		$\theta_{23}^q$ $/^\circ$ &2.38& 2.32 $\to$ 2.44 && $2.37$  \\	
		$\delta^q$ $/^\circ$ & 68.75 & 64.25 $\to$ 73.25  & & $60.25$\\
		$m_u$ $/ \mathrm{MeV}$ & 1.28 & 0.76$\to$ 1.55 && $1.29$  \\	
		$m_c$ $/ \mathrm{GeV}$ & 0.626 & 0.607 $\to$ 0.645 &&  $0.626$ \\	
		$m_t$ $/\mathrm{GeV}$  	  & 171.6& 170 $\to$ 173  && $171.6$ \\
		$m_d$ $/ \mathrm{MeV}$ & 2.74 & 2.57 $\to$ 3.15 &&  $2.75$ \\	
		$m_s$ $/ \mathrm{MeV}$ & 54 & 51 $\to$ 57 && $51$ \\
		$m_b$ $/ \mathrm{GeV}$	  & 2.85 &  2.83 $\to$  
	2.88 &&  $2.91$\\
		\hline
		$\chi^2$ & & & & $12.4$ \\
		\hline		
	\end{tabular}
		\caption{Best fit parameter values for the model.} 
	\label{tab:fit}
\end{table}

We make use of the Mathematica Mixing Parameter Tools package~\cite{Antusch:2005gp} to obtain the flavour observables from the mass matrices in 
Eq.(\ref{eq:massmat}), and then minimize the $\chi^2$ function for the 15 free parameters.
For consistency with the golden relation, all quark and charged lepton masses run to the same energy scale, which we choose to be $M_Z$ \cite{Antusch:2013jca}.
Since the effect on CKM and neutrino parameters induced by the running to $M_Z$ is negligible \cite{Xing:2007fb,Antusch:2013jca}, neutrino oscillation parameters are taken from the global fit \cite{deSalas:2017kay}, 
while the rest of the observables from the PDG \cite{Tanabashi:2018oca}.
The results of our flavour fit are summarized in Table \ref{tab:fit}. 
The minimum happens to be at $\chi^2\approx 12$, making the model remarkably realistic, both in terms of mass as well as mixing predictions.
In what follows we go through the main physics predictions in more detail.  
 
\subsection{Golden quark-lepton mass relation}
\label{sec:golden-relation}

\begin{figure}[h]
\includegraphics[width=0.6\textwidth]{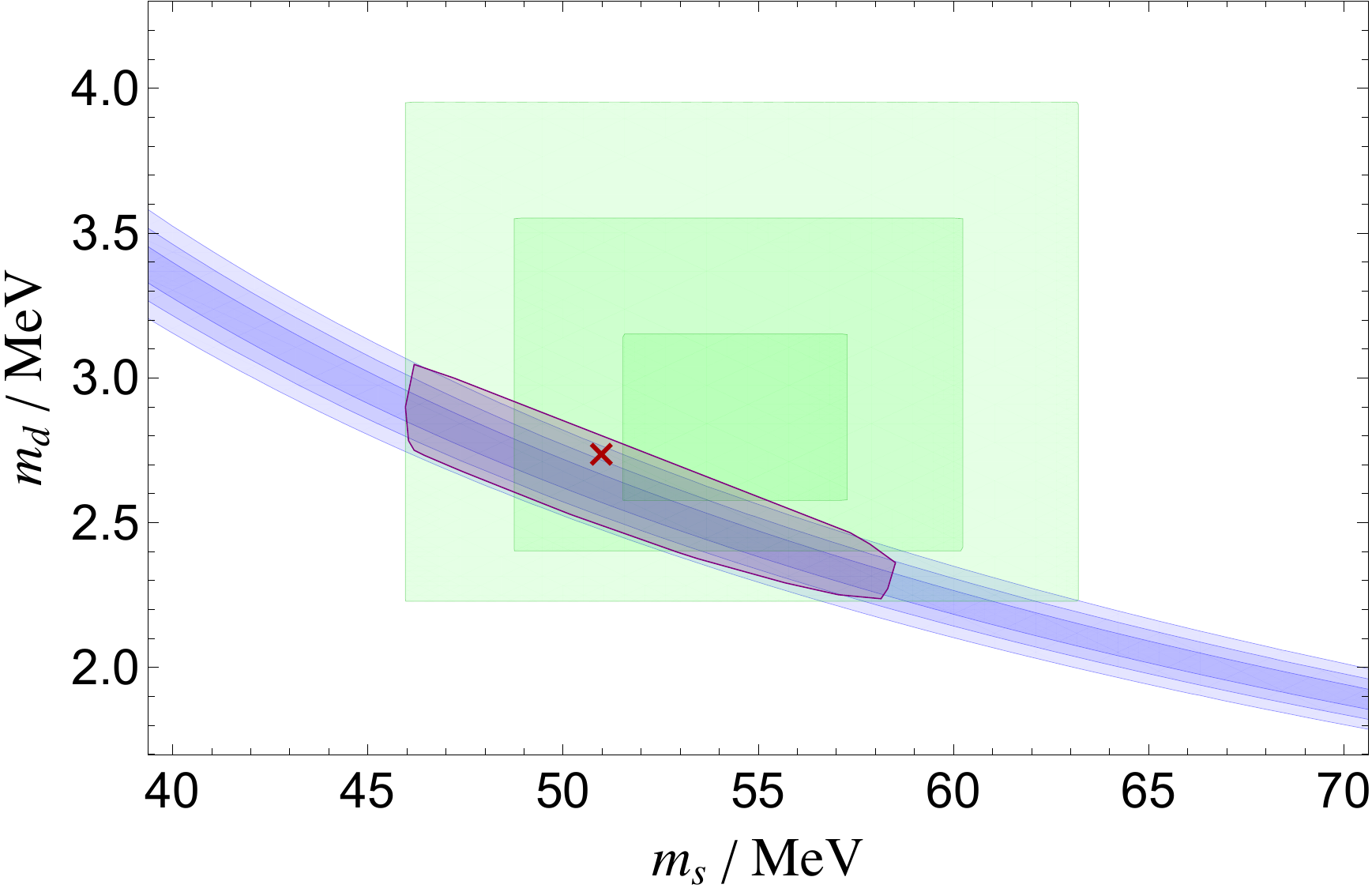}
\caption{Prediction for the down and strange-quark masses at the $M_Z$ scale.
  The blue shades correspond to the 1,  2 and 3$\sigma$ allowed regions from the predicted golden relation $m_\tau/\sqrt{m_\mu m_e}=m_b/\sqrt{m_s m_d}$.
  The green shades indicate the 1, 2 and 3$\sigma$ ranges of measured quark masses, from~\cite{Antusch:2013jca}.
  The purple region is the one consistent at 3$\sigma$ with the global flavour fit in Table \ref{tab:fit}. The red cross indicates the location of the best fit point.} 
\label{fig:golden}
\end{figure}

The purple shaded bands in Figure \ref{fig:golden} show the model predictions for the down and strange-quark masses.
We indicate the 1, 2 and 3$\sigma$ regions extracted from the exact golden relation $m_\tau/\sqrt{m_\mu m_e}=m_b/\sqrt{m_s m_d}$ at the same scale.
These should be compared with the corresponding 1,  2 and 3$\sigma$ allowed quark mass values at $M_Z$ (green shades).
  For this analysis, we have explored the full range of values of the model parameters consistent at 3$\sigma$ with all the 19 measured physical parameters.
  One sees that it largely coincides with the overlap of the green and blue regions. For completeness, the best fit point is indicated with a red cross.

Notice that the golden quark-lepton mass relation does not require the implementation of any unification symmetry to relate quarks and leptons at a fundamental level.
  It emerges exclusively from the properties of the remnant $A_4$ symmetry. However, from the fit in Table \ref{tab:fit}, one can see that the relations 
  \begin{equation}
y_2^d/y_1^e\approx 3/2,\ \ \ y_1^d/y_2^e\approx (3/2)^{-2},
\end{equation}
hold with good precision, suggesting a viable embedding of the present model into a genuine Grand Unification scenario, which would correlate the Yukawa couplings $y^d_{1,2}\sim y^e_{2,1}$.
In particular, the above relations can be exactly realized in a SU(5) model with an adjoint VEV \cite{Antusch:2009gu,Bjorkeroth:2015ora}.
In this way the golden relation can be supplemented with further unification restrictions on the possible values of the Yukawa couplings, and the charged lepton masses can become completely determined from the down-type quark ones.
\subsection{Neutrinoless Double Beta Decay}
\label{sec:neutr-double-beta}

Concerning the neutrino mass spectrum, the analysis in Table \ref{tab:fit} shows that the best fit point displays Normal Ordering (NO) for the neutrino masses, and a rather small neutrino absolute scale $m_1=4.12\,\mathrm{meV}$. 
On closer inspection, by allowing the model parameters to take values compatible at 3$\sigma$ with all the measured observables, one can show that NO always emerges and the resulting lightest neutrino mass lies in the range $m_1=(3.28 - 5.41)\,\mathrm{meV}$.\\[-.2cm]

\begin{figure}[h]
\includegraphics[width=0.6\textwidth]{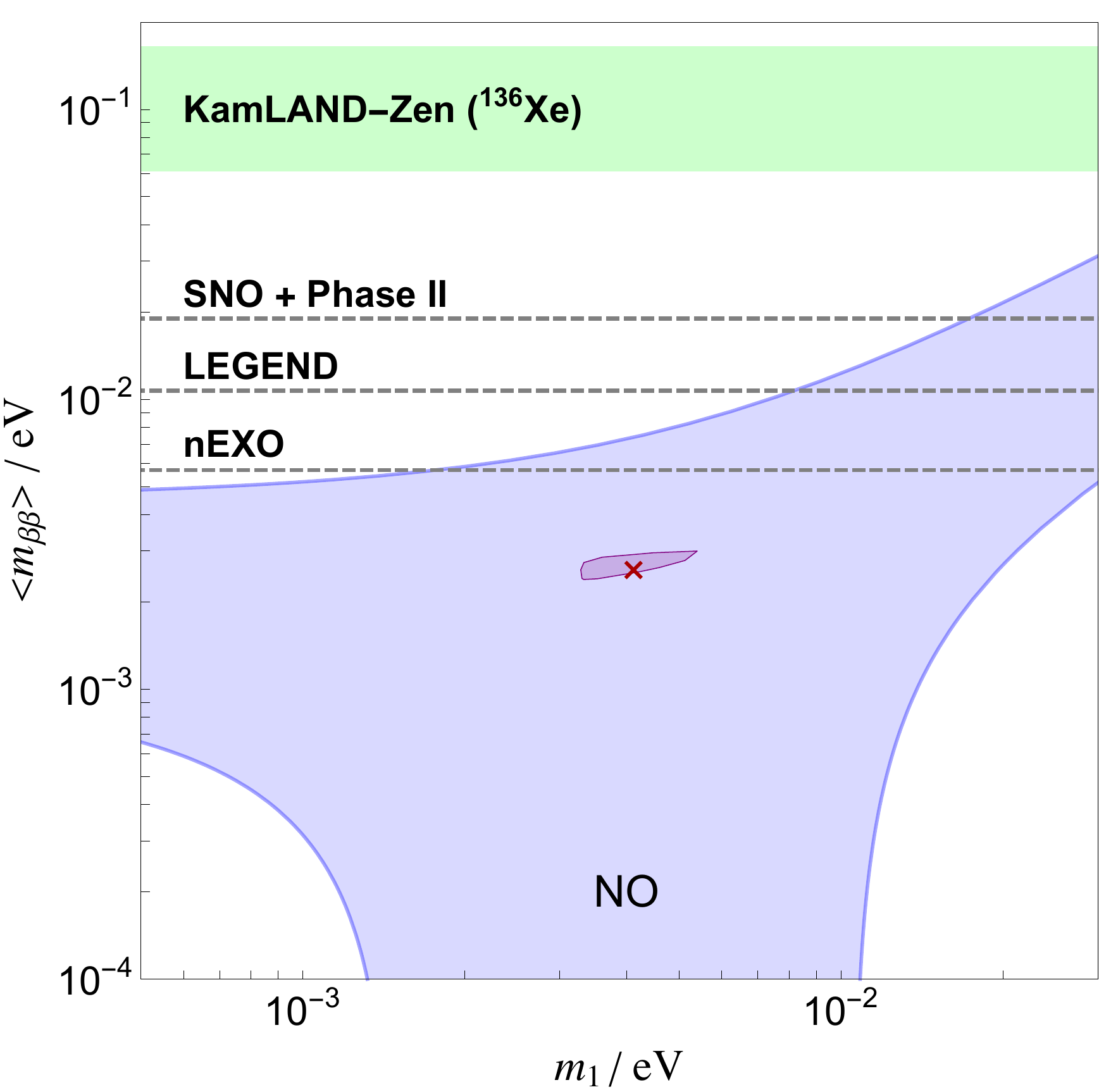}
\caption{Effective Majorana neutrino mass parameter $\vev{ m_{\beta\beta}}$ as a function of the lightest active neutrino mass $m_{1}$.
  The mass spectrum is normal ordered. The blue region is the generic one consistent with oscillations at 2$\sigma$.
  The tiny predicted purple region is the one allowed at 3$\sigma$ in the global fit of all the measured observables in Table \ref{tab:fit}, with the red cross indicating the best fit point.
  The current KamLAND-Zen limit is shown in green, and the projected sensitivities are indicated in dashed horizontal lines, see text.} 
\label{mee}
\end{figure}

Since all neutrino mass parameters as well as Majorana phases are fixed, the model makes concrete predictions for the neutrinoless double beta decay rates.
The resulting effective neutrino mass for the best fit point is given by
\begin{equation}
\vev{ m_{\beta\beta}}=2.65\,\mathrm{meV}.
\end{equation}
A more detailed analysis is depicted in Figure \ref{mee}, where we plot the predicted values for $\vev{ m_{\beta\beta}}$ obtained by randomly varying the model parameters within a range that covers all measured observables inside the $3\sigma$ experimentally allowed range.
One sees that $\vev{ m_{\beta\beta}}$ is determined within rather small errors, and presents a challenge for the next generation of \znbb searches.
For comparison, the top green horizontal band in Figure \ref{mee} represents the current experimental limits from Kamland-Zen $(61 - 165\; \mathrm{meV})$ \cite{KamLAND-Zen:2016pfg}, while the dashed horizontal lines correspond to the projected most optimistic sensitivities from LEGEND $(10.7 - 22.8\; \mathrm{meV})$ \cite{Abgrall:2017syy}, SNO + Phase II $(19 - 46 \; \mathrm{meV})$ \cite{Andringa:2015tza}, and nEXO $(5.7 - 17.7\; \mathrm{meV})$ \cite{Albert:2017hjq}. 

\subsection {Leptonic Dirac CP phase}
\label{sec:leptonic-cp-phase}

\begin{figure}[h]
\includegraphics[width=0.6\textwidth]{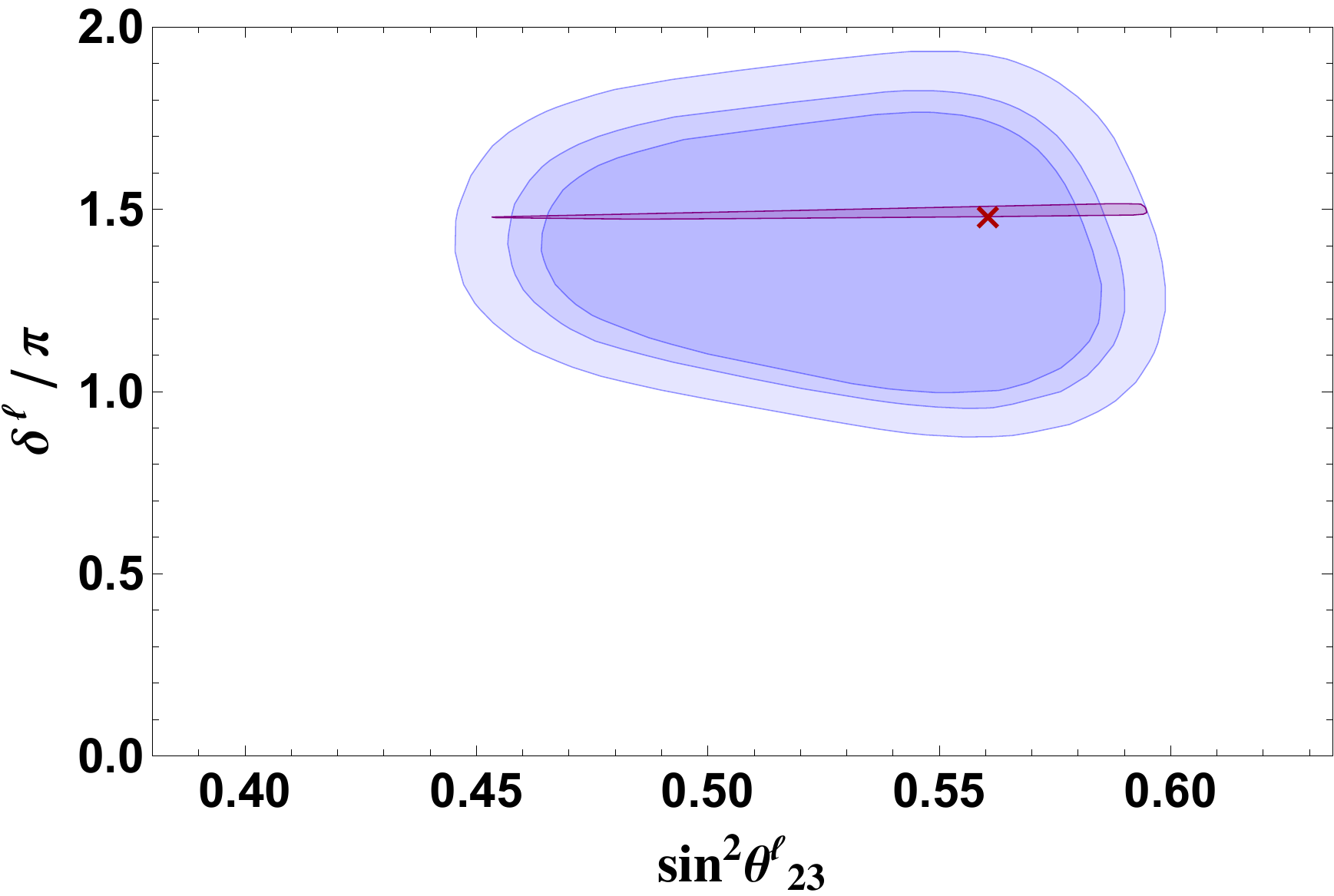}
\caption{Allowed values for the atmospheric angle and the Dirac phase $\delta^\ell$.
  The sharply predicted purple region is compatible at 3$\sigma$ with all the flavour observables,
  while the blue shades represent the 90, 95 and 99\% C.L. regions from the Global oscillation fit in \cite{deSalas:2017kay}. The best fit point is indicated by a red cross.} 
\label{deltavs23} 
\end{figure}

To conclude this section, we show in Figure \ref{deltavs23} the model prediction for the Dirac phase in the lepton sector $\delta^\ell$.
This is the phase affecting neutrino oscillations, and is displayed against the atmospheric angle $\theta^\ell_{23}$.
 These parameters are the target of the next generation of long baseline oscillation experiments. In this plot, the thin purple region corresponds to the model solutions compatible at 3$\sigma$ with all the measured flavour observables, 
while the blue shades represent the 90, 95 and 99\% C.L. regions from the generic oscillation Global fit in \cite{deSalas:2017kay}. 

Here one can see the sharp prediction of the leptonic Dirac phase around $\delta^\ell\approx 3\pi/2$, which arises from the common origin of all CP violating phases \cv{from a single phase $\omega$ at high energies}, 
and the fact that there are no more available phases in the model. 
The later would allow greater deviations from the central values shown in Table \ref{tab:fit}. 
By contrast, one sees this model does not improve the determination of the atmospheric-angle problem, since it can take values over almost all its allowed range of $\theta^{\ell}_{23}$. 
Notice, however, that the atmospheric-angle best fit value preferred by the model lies in the higher octant.

In short, all the results of our flavour fit lie within the 2$\sigma$ region of their measured  values.
The main deviation happens in $\delta^{q}$ and can be easily corrected by relaxing the requirement of real VEVs in Eq.(\ref{vevs}),
by allowing the Higgs fields to develop arbitrary complex VEVs. 
After discarding unphysical phases, this procedure would only introduce two physical phases into the fermion masses, leading to a better $\chi^2$ fit.
Nevertheless, we have opted to keep the model as minimal and predictive as possible, so as to have all phases $\delta^{q}$ and $\delta^{\ell}$ predicted with good precision from a single fixed geometric phase in the model, namely $\omega=e^{2i\pi/3}$. 
This leads to specially sharp predictions for the lightest neutrino mass and the amplitude for neutrinoless double beta decay, Fig.~\ref{mee}.

\section{Summary and conclusions}
\label{sec:Conclusions}

In this work we have proposed an $A_4$ theory for fermion masses and mixings in which the family symmetry emerges naturally from a six-dimensional spacetime after orbifold compactification.
The $A_4$ remnant flavour symmetry leads to the successful ``golden'' quark-lepton unification formula in Eq.~(\ref{golden}) and Fig.~\ref{fig:golden}.
In the simplest scenario, all quark and lepton CP phases are fixed, {\it i.e.} the four CP violating phases $\delta^q$, and $\phi_{12,13,23}$ all emerge from a single phase induced by the orbifold compactification.
Under these assumptions we have performed a global flavour fit that yields all parameters within their measured 2$\sigma$ ranges.
The leptonic CP phase of neutrino oscillations is sharply predicted as given in Fig.~\ref{deltavs23}.
The resulting lightest neutrino mass lies in the range $m_1=(3.28 - 5.41)\,\mathrm{meV}$. 
The theory gives a sharp prediction for the neutrinoless double beta decay, characterized by an effective mass parameter $\vev{ m_{\beta\beta}} \approx 2.65\,\mathrm{meV}$, Fig.~\ref{mee}.
Some ``tension'' occurs for the quark CP phase $\delta^{q}$ and can be easily fixed 
by allowing the Higgs fields to develop arbitrary complex VEVs. 
This procedure would introduce two physical CP phases, leading to a better global fit of the flavour parameters. 
Finally, we note that the neutrino mass spectrum is normal-ordered, with the best fit value of $\theta^{\ell}_{23}$ at the higher octant, as currently preferred by neutrino oscillation data.

\acknowledgements 
\noindent

Work supported by the Spanish grants SEV-2014-0398 and FPA2017-85216-P (AEI/FEDER, UE), PROMETEO/2018/165 (Generalitat Valenciana) and the Spanish Red Consolider MultiDark FPA2017-90566-REDC.  CAV-A is supported by the Mexican C\'atedras CONACYT project 749 and SNI 58928.

\appendix
\section{$A_4$ basis}
\label{app:a4}

The $A_4$ group can be defined by the presentation
\begin{equation}
A_4\simeq \{S,T|S^2=T^3=(ST)^2=1\}.
\end{equation}
It has 4 irreducible representations, which in the Ma-Rajasekaran basis transform as
\begin{equation}\begin{split}
\textbf{1}&:\ \ \ \ S=1,\ \ \ \ T=1,\\
\textbf{1}'&:\ \ \ \ S=1,\ \ \ \ T=\omega,\\
\textbf{1}''&:\ \ \ \ S=1,\ \ \ \ T=\omega^2,\\
\textbf{3}&:\ \ \ \ S=\left(\begin{array}{ccc} 1&0&0\\ 0&-1&0\\ 0&0&-1
\end{array}\right),\ \ \ \ T=\left(\begin{array}{ccc} 0&1&0\\ 0&0&1\\ 1&0&0
\end{array}\right),\\
\end{split}\end{equation}
where $\omega=e^{2i\pi/3}$. These define the invariant non trivial contractions as
\begin{equation}\begin{split}
\textbf{1}'\times\textbf{1}'=&\textbf{1}'',\ \ \ \textbf{1}''\times\textbf{1}''=\textbf{1}',\ \ \ \textbf{1}''\times\textbf{1}'=\textbf{1},\\
\textbf{3}&\times \textbf{3} =\textbf{1}+\textbf{1}'+\textbf{1}''+\textbf{3}_1+\textbf{3}_2.
\end{split}\end{equation}
The contractions of two triplets $\textbf{3}_a\sim (a_1,a_2,a_3)$ and $\textbf{3}_b\sim (b_1,b_2,b_3)$ are decomposed as
\begin{equation}\begin{split}
\textbf{3}_a\times \textbf{3}_b\to \textbf{1}&=a_1b_1+a_2b_2+a_3b_3,\\\textbf{3}_a\times \textbf{3}_b\to \textbf{1}'&=a_1b_1+\omega^2 a_2b_2+\omega a_3b_3,\\
\textbf{3}_a\times \textbf{3}_b\to \textbf{1}''&=a_1b_1+\omega a_2b_2+\omega^2 a_3b_3,\\
\textbf{3}_a\times \textbf{3}_b\to \textbf{3}_1&=(a_2b_3,a_3b_1,a_1b_2),\\
\textbf{3}_a\times \textbf{3}_b\to \textbf{3}_2&=(a_3b_2,a_1b_3,a_2b_1).\\
\end{split}\end{equation}


\centerline {\bf  Generalized CP symmetry}

\vskip .2cm

We assume a generalized CP symmetry that does not commute with $A_4$:
\begin{equation}
\mathcal{CP}:\ \ \phi_{\textbf{1}}\to\phi_{\textbf{1}}^\dagger,\ \ \phi_{\textbf{1}'}\to\phi_{\textbf{1}''}^\dagger,\ \ \phi_{\textbf{1}''}\to \phi_{\textbf{1}'}^\dagger,\ \ \phi_{\textbf{3}}\to \left(\begin{array}{ccc}1&0&0\\0&0&1\\0&1&0\end{array}\right) \phi_{\textbf{3}}^\dagger,
\end{equation}
which limits the structure of the lagrangian. In particular we study how this transformation affects the various triplet ($\textbf{a},\textbf{b},\textbf{c}$) contractions into singlets
\begin{equation}\begin{split}
\mathcal{CP}(\textbf{a}\textbf{b})_\textbf{1}&=(a_1b_1+a_3b_3+a_2b_2)^*=a^*_1b^*_1+a_3^*b_3^*+a_2^*b_2^*=(\textbf{a}^*\textbf{b}^*)_\textbf{1},\\
\mathcal{CP}(\textbf{a}\textbf{b})_{\textbf{1}'}&=(a_1b_1+\omega^2 a_3b_3+\omega a_2b_2)^*=a^*_1b^*_1+\omega a_3^*b_3^*+\omega^2 a_2^*b_2^*=(\textbf{a}^*\textbf{b}^*)_{\textbf{1}''},\\
\mathcal{CP}(\textbf{a}\textbf{b})_{\textbf{1}'}&=(a_1b_1+\omega a_3b_3+\omega^2 a_2b_2)^*=a^*_1b^*_1+\omega^2 a_3^*b_3^*+\omega a_2^*b_2^*=(\textbf{a}^*\textbf{b}^*)_{\textbf{1}''},\\
\mathcal{CP}(\textbf{a}\textbf{b}\textbf{c})_{\textbf{1}_1}&=(c_1 b_3 a_2+c_3 b_2 a_1+c_2b_1a_3)^*=(\textbf{a}^*\textbf{b}^*\textbf{c}^*)_{\textbf{1}_1}\ ,\\
\mathcal{CP}(\textbf{a}\textbf{b}\textbf{c})_{\textbf{1}_2}&=(c_1b_2 a_3+c_3 b_1a_2+c_2b_3a_1)^*=(\textbf{a}^*\textbf{b}^*\textbf{c}^*)_{\textbf{1}_2}\ ,\\
\end{split}\end{equation}
which are all triplet contractions used in our model. We can conclude that assuming this generalized CP symmetry in this basis is equivalent to just complex conjugating the fields and therefore it implies that the Yukawa couplings \jv{must be} real.

\bibliographystyle{utphys}
\bibliography{bibliography}
\end{document}